\documentclass[aps,prb,twocolumn,groupedaddress,showpacs]{revtex4}

\usepackage{graphicx}
\usepackage{dcolumn}
\usepackage{bm}
\begin{document}

\title{Examination of the $c$-axis resistivity of 
Bi$_{2}$Sr$_{2-x}$La$_x$CuO$_{6+\delta}$ in magnetic fields\\
up to 58 T}

\author{S. Ono}
\author{Yoichi Ando}
\affiliation{Central Research Institute of Electric Power Industry, Komae, 
Tokyo 201-8511, Japan}

\author{F. F. Balakirev}
\author{J. B. Betts}
\author{G. S. Boebinger}
\altaffiliation{Present address: National High Magnetic Field Laboratory, 
Tallahassee, FL 32310, USA} 
\affiliation{NHMFL, Los Alamos National Laboratory, Los Alamos, 
New Mexico 87545, USA}

\date{\today}

\begin{abstract}

We measure the magnetic-field dependence of the $c$-axis resistivity,
$\rho_c(H)$, in a series of Bi$_{2}$Sr$_{2-x}$La$_x$CuO$_{6+\delta}$
(BSLCO) single crystals for a wide range of doping using pulsed magnetic
fields up to 58 T. The behavior of $\rho_c(H)$ is examined in light of
the recent determination of the upper critical field $H_{c2}$ for this
material using Nernst effect measurements. We find that the peak in
$\rho_c(H)$ shows up at a field $H_p$ that is much lower than $H_{c2}$
and there is no discernable feature in $\rho_c(H)$ at $H_{c2}$.
Intriguingly, $H_p$ shows a doping dependence similar to that of $T_c$,
and there is an approximate relation $k_{B}T_c \simeq
\frac{1}{2}g\mu_{B}H_p$. Moreover, we show that the data for the
lowest-$T_c$ sample can be used to estimate the pseudogap closing field
$H_{pg}$, but the method to estimate $H_{pg}$ proposed by Shibauchi {\it
et al.} [Phys. Rev. Lett. {\bf 86}, 5763 (2001)] must be modified to
apply to the BSLCO system.

\end{abstract}

\pacs{74.25.Fy, 74.25.Dw, 74.72.Hs}

\maketitle

\section{INTRODUCTION}

In high-$T_c$ cuprates, the $c$-axis transport occurs as a tunneling
process, and therefore signifies the density of electrons available for
the tunneling as well as the tunneling matrix elements. \cite{Gray} As a
result, the $c$-axis resistivity $\rho_c$ is a useful probe
\cite{Lavrov,Ono} of such features as the pseudogap \cite{Timusk} or the
superconducting correlations \cite{Millis} above $T_c$. On the other
hand, there are a number of open questions regarding the interpretation
of the magnetic-field ($H$) dependence of $\rho_c$ below $T_c$, in which
the suppression of superconductivity and the subsequent negative
magnetoresistance (MR) at higher $H$ defines a peak value of $\rho_c$ at
$H_p$. One question is whether the magnetic-field region above $H_p$ can
be viewed as the normal state and, if not, how one can determine the
upper critical field $H_{c2}$. \cite{AndoHc2,Morozov} Another question
is whether the $\rho_c(H)$ data can be used to derive a characteristic
field for the closing of the pseudogap by the Zeeman splitting.
\cite{Shibauchi,KrEl1,KrEl2}

It was argued by Morozov {\it et al.} \cite{Morozov} that $H_p$
separates the two regions in the superconducting state, one dominated by
Cooper pair tunneling and the other dominated by quasiparticle
tunneling. This proposal has been backed up by more recent argument
\cite{KrEl2} and it seems indeed likely that $H_p$ signifies a crossover
from a phase-coherent regime (where the $c$-axis transport is dominated
by the Cooper pair tunneling) to a phase-incoherent regime. In this
sense, if one assumes that the phase coherence is the defining factor
for the superconducting state in cuprates, one can identify that $H_p$
is the characteristic field for superconductivity, although it clearly
lies below the mean-field $H_{c2}$ which describes the onset of
superconducting pair correlations. (Therefore, whether to call the
region between $H_p$ and $H_{c2}$ the ``normal state" is a matter of
semantics; ``fully resistive state" might better suit this regime that
is so strikingly different from the normal state of BCS
superconductors.) 

Later, Shibauchi {\it et al.} argued \cite{Shibauchi} that the negative
MR data above $H_p$ can be used to estimate the field at which the
pseudogap collapses due to the increasing Zeeman energy, calling this
field the pseudogap closing field $H_{pg}$. Although their procedure
relies on determining the putative intrinsic $\rho_c$ in the absence of
the pseudogap and a necessary extrapolation to determine a value for
$H_{pg}$, their central assertion is that the negative MR comes from a
recovery of the electronic density of states near the Fermi energy $E_F$
that is suppressed in the pseudogap state. The work by Shibauchi {\it et
al.} \cite{Shibauchi,KrEl1} was done on
Bi$_{2}$Sr$_{2}$CaCu$_{2}$O$_{8+\delta}$ (Bi-2212) for which the
intrinsically high $T_c$ makes the measurements and the analysis
inherently difficult; it would be useful to examine $H_{pg}$ in another
cuprate that has lower $T_c$ and thus is expected to have lower
characteristic magnetic-field scales. From this point of view, the
Bi$_{2}$Sr$_{2-x}$La$_x$CuO$_{6+\delta}$ (BSLCO) system is particularly
suitable for examining the behavior of $\rho_c(H)$, because the $T_c$ of
this system never exceeds 40 K and one can obtain high-quality single
crystal samples for a wide range of hole doping. \cite{Murayama,Ono}

Recently, it was shown that the Nernst effect in cuprates is a useful
probe of the presence of vortices and, hence, superconducting
correlations, \cite{WangNature} from which Wang {\it et al.} deduced the
pseudogap onset temperature above $T_c$ (Ref. \onlinecite{WangPG}) and
$H_{c2}$ below $T_c$ (Ref. \onlinecite{WangHc2}). In particular, recent
Nernst effect measurements in magnetic fields up to 45 T make a very
good case \cite{WangHc2} that the vortex Nernst signal disappears above
a well-defined field $H_{c2}^N$ and it is reasonable to consider that
$H_{c2}^N$ marks the field where the superconducting pair correlations
disappear, {\it i.e.}, the upper critical field. Therefore, it would be
illuminating to compare the information obtained by $\rho_c$
measurements with that obtained by Nernst effect measurements. The BSLCO
system is ideal for this purpose as well, because detailed Nernst effect
measurements have already been performed on BSLCO.
\cite{WangHc2,WangNew}

In this work, we measure $\rho_c$ of a series of high-quality BSLCO
single crystals in pulsed magnetic fields up to 58 T and examine the
implication of the observed $\rho_c(H)$ behavior in the context of
Nernst effect measurements that were performed on the samples from the
same batch. It is found that the doping dependence of $H_p$ essentially
tracks that of $T_c$, and, moreover, there is an approximate relation
$1.3T_c$ (in Kelvin) $\simeq H_p$ (in Tesla), which suggests that the
electronic Zeeman energy at $H_p$ ($\frac{1}{2}g\mu_{B}H_p$) equals the
thermal energy $k_{B}T_c$. Also, our $\rho_c(H)$ data are featureless at
$H_{c2}^N$ ($H_{c2}$ as determined by the Nernst signal), which
demonstrates that it is not possible to determine $H_{c2}$ from current
state-of-the-art resistivity experiments using pulsed magnetic fields.
Furthermore, our data support the definition of a pseudogap closing
field $H_{pg}$ which can in principle be deduced from $\rho_c(H)$
behavior; however, we find that the procedure employed by Shibauchi {\it
et al.} \cite{Shibauchi} is not appropriate to correctly obtain
$H_{pg}$.

\begin{table}
\caption{Actual hole concentrations per Cu, $p$, 
the zero-resistivity temperature $T_0$, and the peak temperature $T_p$ 
(which marks the onset of the superconducting transition) for each La 
concentration $x$.  The $p$ values are determined 
from the empirical relation between $x$ and $p$ obtained in Ref. 
\onlinecite{Hanaki}.}
\begin{ruledtabular}
\begin{tabular}{cccccc}
 $x$& 0.23& 0.39& 0.49& 0.66& 0.84\\
\hline
$p$& 0.18& 0.16& 0.14& 0.12& 0.10\\
$T_0$& 22& 32& 28& 26& 4\\
$T_p$& 25& 34& 30& 28& 8\\
\end{tabular}
\end{ruledtabular}
\end{table}

\section{EXPERIMENTS}

The Bi$_{2}$Sr$_{2-x}$La$_x$CuO$_{6+\delta}$ (BSLCO) crystals used for
this study are grown by the floating-zone method \cite{Ono} and they are
the same as the ones used in our recent study of the
$\rho_c(H)/\rho_{ab}(H)$ resistivity anisotropy in the fully resistive
state. \cite{QPT} We note that the series of BSLCO samples used in the
recent Nernst effect measurements by Wang {\it et al.}
\cite{WangPG,WantHc2,WangNew} are obtained from the same batches. In the
present study, to corroborate the data for the La-doped samples, we also
measure one La-free sample with the composition of
Bi$_{2.13}$Sr$_{1.89}$CuO$_{6+\delta}$ (denoted ``La-free"), which shows
zero resistivity at 9.1 K. For all the La-doped samples, we list in
Table I the actual La content $x$, the corresponding \cite{Hanaki}
doping concentration per Cu, $p$, and the zero-resistivity temperature
$T_0$, as well as the peak temperature in the $\rho_c(T)$ curves, $T_p$.
All the crystals are annealed according to the recipe described in our
previous paper \cite{Ono} to optimize the sharpness of the
superconducting transition. 

The samples for the $\rho_c$ measurements are prepared by hand-painting
ring-shaped current contacts and small circular voltage contacts in the
center of the current-contact ring on the opposing $ab$ faces of the
crystals. \cite{Ono} The $\rho_c(H)$ data are measured at fixed
temperatures using a high-frequency ($\sim$100 kHz) four-probe technique
\cite{logT,MI,OnoMI} during the 15 msec duration of the 58-T pulsed
magnetic fields. As always, we pay particular attention to make sure
that the data are not adversely affected by eddy-current heating.
\cite{logT,OnoMI} The temperature dependences of $\rho_c$ of the present
samples in zero magnetic field are essentially the same as those we
reported previously. \cite{Ono}

\begin{figure} 
\includegraphics[width=8.5cm]{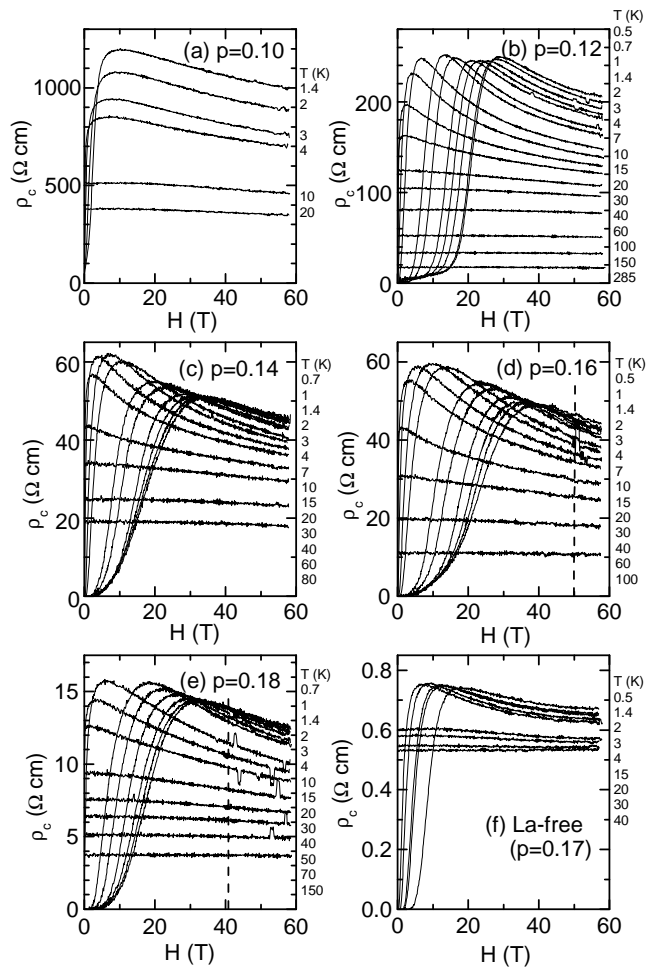}
\caption{Magnetic-field dependence of $\rho_c$ at selected temperatures
in BSLCO for a wide range of doping: (a) $p$ = 0.10, (b) $p$ = 0.12, 
(c) $p$ = 0.14, (d) $p$ = 0.16, (e) $p$ = 0.18, and (f) La-free ($p$ =
0.17). The position of $H_{c2}^N$ is marked by a vertical dashed line. 
$H_{c2}^N$ data is determined in Ref. \onlinecite{WangHc2}.} 
\end{figure}

\section{RESULTS AND DISCUSSIONS}

Figure 1 shows the $\rho_c(H)$ curves at various temperatures for all six
samples studied. From these data, we determine $H_p(T)$ for all the
samples and plot them in Fig. 2(a). Similarly to Bi-2212, \cite{Morozov}
$H_p(T)$ of all the samples (except for $p$ = 0.10) shows a pronounced
upward curvature and steeply increases at low temperature. However,
these temperature dependences are not really diverging, and one can
obtain a reasonable fit to the data with an exponential function
\cite{KrEl2} $H_p = H_{p0} \exp(-T/T_0)$ for the low-temperature part;
such fit gives an estimate of $H_p$ in the zero-temperature limit,
$H_{p0}$. Figure 2(b) shows the doping dependence of $H_{p0}$, together
with the measured $H_p$ values at 1.4 K and 4 K. It is clear that these
doping dependences are similar to that of $T_c$. To make a meaningful
comparison, we consider the temperature $T_p$, where $\rho_c(T)$ shows a
peak, to characterize the crossover between quasiparticle-dominated
transport to the Cooper-pair dominated transport, similarly to $H_p$. In
other words, $T_p$ is a measure of the onset $T_c$. The doping
dependence of $T_p$ is also plotted in Fig. 2(b) using the
right-hand-side axis. Intriguingly, $1.3T_p$ (in Kelvin) is roughly
equal to $H_p$ (in Tesla), which suggests $k_{B}T_c \simeq
\frac{1}{2}g\mu_{B}H_p$. This means that both the thermal energy at
$T_c$ ($k_{B}T_c$) and the electronic Zeeman energy at
$H_p$($\frac{1}{2}g\mu_{B}H_p$) give the single energy scale required to
destroy the phase coherence. A similar relation has also been reported
for Bi-2212.\cite{KrEl2}

\begin{figure} 
\includegraphics[width=7cm]{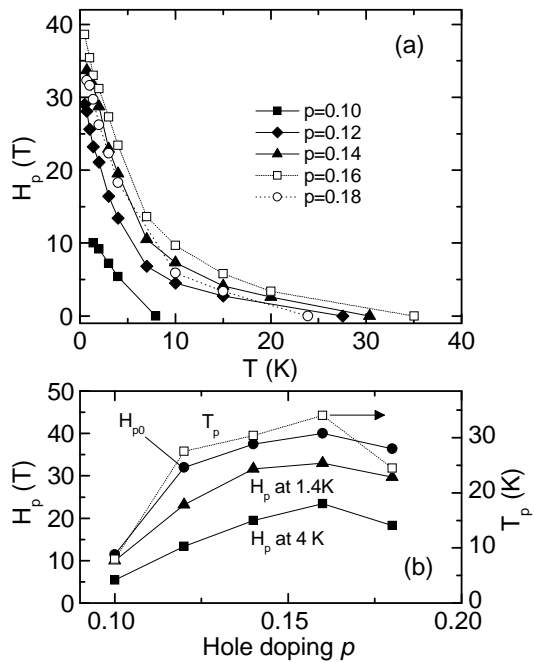}
\caption{(a) Temperature dependences of the peak field $H_p$ for various 
dopings. The lower panel (b) shows the doping dependence of $H_{p0}$ 
(solid circles) as well as the measured $H_p$ values at 1.4 K 
(solid triangle) and 4 K (solid squares). The doping dependence of $T_p$ 
(open squares) is also plotted. }
\end{figure}

Now we compare our result with the Nernst effect
measurements.\cite{WangHc2} Wang {\it et al.} have measured
\cite{WangHc2} the Nernst effect in our BSLCO samples at $p$ = 0.12,
0.16, and 0.18, which corresponds to the La content of 0.6, 0.4, and
0.2, respectively. \cite{note} Their data for $p$ = 0.16 extend to 45 T
and with very little extrapolation give $H_{c2}^N$ of 50 T. This
$H_{c2}^N$ is essentially temperature independent at low temperatures.
For other dopings, Wang {\it et al.} obtained \cite{WangHc2} $H_{c2}^N$
values of 65 and 41 T for $p$ = 0.12 and 0.18, respectively. In Figs.
1(d) and 1(e), the position of $H_{c2}^N$ is marked by a vertical line.
[The $H_{c2}$ value determined for $p$ = 0.12 is above the range of the
present experiment and thus is not shown in Fig. 1(b).] There is no
discernible feature in our $\rho_c(H)$ curves at $H_{c2}^N$, implying
that the onset of superconducting pair correlations does not noticeably
affect $\rho_c$ because $\rho_c$ is dominated by quasiparticle
tunneling. Note that the same situation is known for the in-plane
resistivity $\rho_{ab}$. \cite{Bergemann} Most likely, the extremely
strong phase fluctuations in the cuprates play a key role, allowing the
full recovery of the normal-state resistivity at a magnetic field
smaller than $H_{c2}$. In any case, these data demonstrate that it is
impractical or impossible to deduce $H_{c2}$ from resistivity
measurements.

\begin{figure} 
\includegraphics[width=8cm]{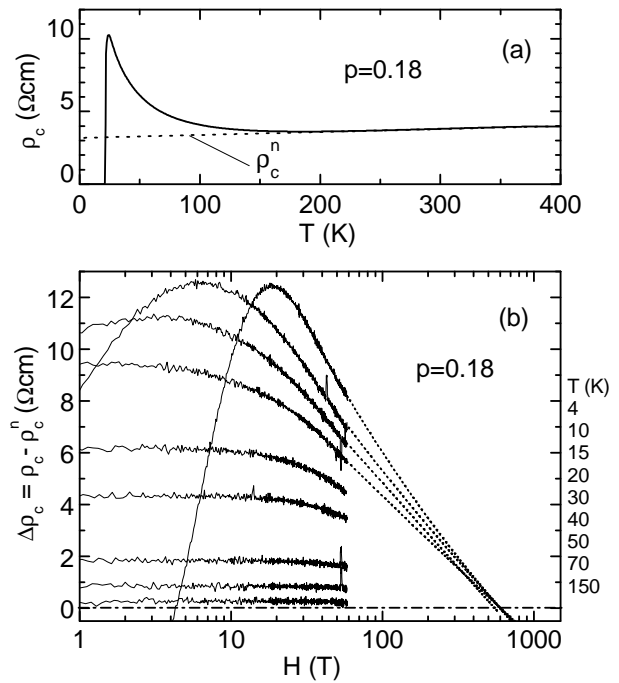}
\caption{(a) Temperature dependence of $\rho_c$ for $p$ = 0.18; the
dashed line is an extrapolation of the high-temperature $\rho_c(T)$ 
to zero temperature, giving the estimate of $\rho_c^n$. 
(b) $\Delta\rho_c$(H) [$\equiv \rho_c(H) - \rho_c^n$] at selected
temperatures for $p$ = 0.18. Dotted lines in (b) are fits of the
high-field data to $\Delta\rho_c(H) = \Delta\rho_c(0) + bH^{\alpha}$ 
and its extrapolation, following the procedure of Shibauchi et al.
\cite{Shibauchi,KrEl1}}
\end{figure}

Next we examine whether the present data for $\rho_c(H)$ can be used to
deduce the pseudogap closing field $H_{pg}$. According to the procedure
proposed by Shibauchi {\it et al.},\cite{Shibauchi} one first determines
the putative $\rho_c$ in the {\it absence} of the pseudogap, $\rho_c^n$,
\cite{note2} by linearly extrapolating the high-temperature part of
$\rho_c(T)$ where it shows a metallic behavior ($d\rho_c/dT > 0$). As
shown in Fig. 3(a), for our overdoped sample ($p$ = 0.18), such an
extrapolation gives $\rho_c^n$ of about 3 m$\Omega$cm at low
temperature. One then calculates $\Delta\rho_c(H) \equiv \rho_c(H) -
\rho_c^n$ and fits the high-field part of $\Delta\rho_c(H)$ with an
empirical formula \cite{Shibauchi,KrEl1} $\Delta\rho_c(H) =
\Delta\rho_c(0) + bH^{\alpha}$; extrapolation of this fit to
$\Delta\rho_c$ = 0 gives the estimate of $H_{pg}$ in the manner of
Shibauchi {\it et al.} When applied to our $p$ = 0.18 data, this
analysis gives an estimate of $H_{pg}$ of about 600 T [see Fig. 3(b)],
which is almost certainly too high for an overdoped sample and suggests
the inapplicability of the procedure proposed by Shibauchi {\it et al.}
for determining $H_{pg}$, at least for the BSLCO system. The reason for
the inapplicability probably lies in the assumptions used to determine
$\rho_c^n$: as we have shown in our previous paper, \cite{Ono} the
``insulating" temperature dependence of $\rho_c$ comes not only from the
pseudogap but also from the charge confinement effect. Because of the
existence of the latter, the assumption of a $T$-linear $\rho_c^n$ down
to the lowest temperature becomes dubious. Therefore, we claim that any
determination of $H_{pg}$ from resistivity data should not rely on any
assumptions about $\rho_c^n$ or $\rho_c(H)$.

\begin{figure} 
\includegraphics[width=7.5cm]{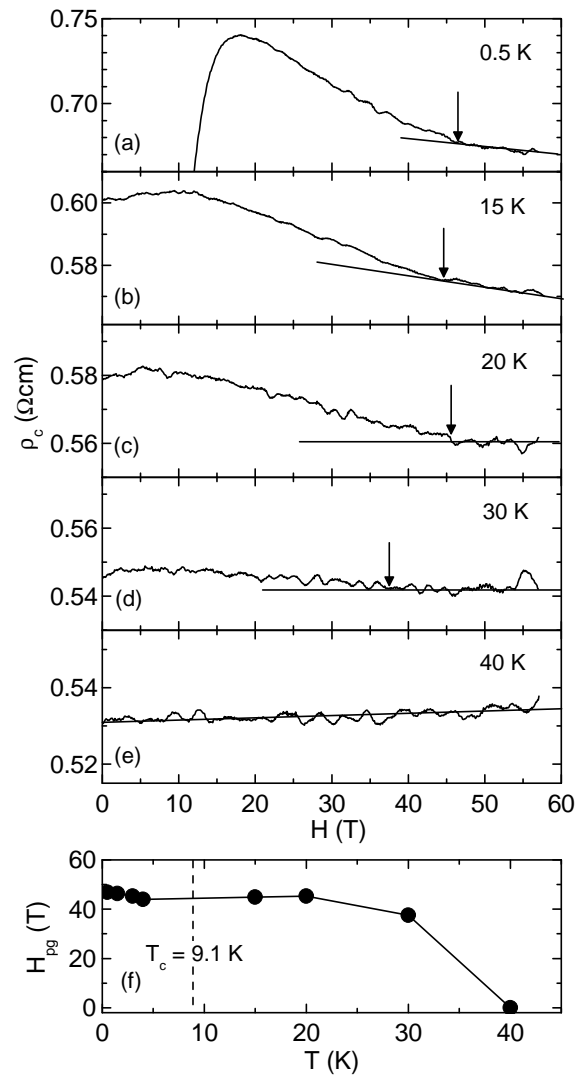}
\caption{(a)-(e) $\rho_c(H)$ of the La-free sample at selected
temperatures; here the data are mildly filtered to remove the
high-frequency noise apparent in the raw data shown in Fig. 1(f). The
solid lines are linear fits of the high-field data. Arrows mark the
field above which the $\rho_c(H)$ shows a near-saturation and thus 
would corresponds to $H_{pg}$. (f) Temperature dependence of $H_{pg}$
obtained from the above method.} 
\end{figure}

Incidentally, the $\rho_c(H)$ data of our La-free sample (whose $p$
value has been estimated \cite{Ono} to be 0.17) shows a behavior that is
almost saturating at high field even at the lowest temperature. This is
probably because this sample has the lowest $T_c$ ($T_0$ = 9.1 K and
$T_p$ = 10.2 K) and accordingly low magnetic field scales. As one can
see in Figs. 4(a)-4(e), the high-field $\rho_c$ is saturating to a value
which increases with decreasing temperature, indicating that the true
$\rho_c^n$ presents an ``insulating" behavior ($d\rho_c/dT < 0$) even
when the pseudogap is closed by the magnetic field. Also, one can
crudely estimate $H_{pg}$ from this near-saturation as shown by the
arrows in Figs. 4(a)-4(d). [The solid straight lines are the fits to the
region where we consider the rapid decrease of $\rho_c$ is finished;
these lines at low temperatures are slightly sloped, which may mean that
there is some intrinsic negative MR in the absence of the pseudogap or
mean that the pseudogap is not yet fully closed.] Intriguingly, while
there is no negative MR (and thus there appears to be no pseudogap) at
40 K, by 30 K the pseudogap opens and the $H_{pg}$ suggested by the data
is already higher than 30 T.

It is useful to note that if one were to apply the same method of
extracting $H_{pg}$ that we demonstrated for the La-free sample to the
data for $p$ = 0.18, the estimated $H_{pg}$ would be larger than 60 T,
because there is no saturation below 60 T [see Fig. 1(e)]. This might
seem rather odd, since the doping level in the La-free sample is $p$ =
0.17, which is slightly more underdoped than $p$ = 0.18, and yet the
estimated $H_{pg}$ for the La-free sample would be smaller than that for
$p$ = 0.18; normally, one would expect $H_{pg}$ to be larger in more
underdoped samples. However, one must take into account the fact that
the $T_c$ of the La-free samples is significantly lower than that of the
La-doped samples at the same doping level, which strongly suggests that
there exists some additional pair-breaking mechanism in the La-free
samples. Remember, as has been argued by Shibauchi {\it et al.},
\cite{Shibauchi} $H_{pg}$ is likely to reflect the spin singlet
formation; thus, if there is an additional pair-breaking mechanism in
the La-free sample, it is rather natural for $H_{pg}$ to become
accordingly small. A recent work by Eisaki {\it et al.} \cite{Eisaki}
reported a clear relationship between $T_c$ and the cation disorder in
the Sr site ($A$-site disorder) for the single-layer Bi-based cuprates,
so that the strong $A$-site disorder caused by excess Bi in the La-free
samples is likely to be responsible for the strong pair breaking.

\section{CONCLUSIONS}

We measure and examine the behavior of $\rho_c(H)$ for a series of BSLCO
samples in magnetic fields up to 58 T. The salient points are: (i) The
peak field in the zero-temperature limit, $H_{p0}$, shows a dome-shaped
doping dependence and is related to $T_c$ via the relation $k_{B}T_c
\simeq \frac{1}{2}g\mu_{B}H_p$, which is understandable if both $T_c$
and $H_{p0}$ are determined by the onset of phase coherence. (ii) There
is no feature in the $\rho_c(H)$ data at the upper critical field
determined by the Nernst effect, $H_{c2}^N$. (iii) The pseudogap closing
field $H_{pg}$ can be determined by $\rho_c(H)$ in overdoped samples
with low $T_c$, but one should not employ an extrapolation of
high-temperature $\rho_c(T)$ to low temperatures in its determination,
because one cannot {\it a priori} know the temperature dependence of the
$c$-axis resistivity in the absence of the pseudogap.

\begin{acknowledgments}
The work at CRIEPI was supported in part by the Grant-in-Aid for Science
provided by the Japanese Society for the Promotion of Science. The NHMFL
is supported by the NSF, the State of Florida and the DOE.
\end{acknowledgments}



\begin{thebibliography}{99}

\bibitem{Gray}
For a review, see S. L. Cooper and K. E. Gray, in 
\textit{Physical Properties of High Temperature Superconductors IV}, 
edited by D. M. Ginsberg (World Scientific, Singapore, 1994).

\bibitem{Lavrov}
A. N. Lavrov, Y. Ando, and S. Ono, Europhys. Lett. \textbf{57}, 267 (2002).

\bibitem{Ono}
S. Ono and Y. Ando, Phys. Rev. B \textbf{67}, 104512 (2003).

\bibitem{Timusk}
T. Timusk and B. Statt, Rep. Prog. Phys. \textbf{62}, 61 (1999).

\bibitem{Millis}
J. Orenstein and A. J. Millis, Science \textbf{288}, 468 (1998).

\bibitem{AndoHc2}
Y. Ando, G. S. Boebinger, A. Passner, L. F. Schneemeyer, T. Kimura, M. Okuya, 
S. Watauchi, J. Shimoyama, K. Kishio, K. Tamasaku, N. Ichikawa, and S. Uchida, 
Phys. Rev. B \textbf{60}, 12475 (1999).

\bibitem{Morozov}
N. Morozov, L. Krusin-Elbaum, T. Shibauchi, L. N. Bulaevskii, M. P. Maley, 
Yu. I. Latyshev, and T. Yamashita, Phys. Rev. Lett. \textbf{84}, 1784 (2000).

\bibitem{Shibauchi}
T. Shibauchi, L. Krusin-Elbaum, M. Li, M. P. Maley, and P. H. Kes, 
Phys. Rev. Lett. \textbf{86}, 5763 (2001).

\bibitem{KrEl1}
L. Krusin-Elbaum, T. Shibauchi, and C. H. Mielke, 
Phys. Rev. Lett. \textbf{92}, 097005 (2004).

\bibitem{KrEl2}
L. Krusin-Elbaum, G. Blatter, and T. Shibauchi, 
Phys. Rev. B \textbf{69}, 220506 (2004).

\bibitem{Murayama}
Y. Ando and T. Murayama, Phys. Rev. B \textbf{60}, R6991 (1999).

\bibitem{WangNature}
Z. A. Xu, N. P. Ong, Y. Wang, T. Kakeshita, and S. Uchida, 
Nature \textbf{406}, 489 (2000).

\bibitem{WangPG}
Y. Wang, Z. A. Xu, T. Kakeshita, S. Uchida, S. Ono, Y. Ando, and N. P. Ong, 
Phys. Rev. B \textbf{64}, 224519 (2001).

\bibitem{WangHc2}
Y. Wang, S. Ono, Y. Onose, G. Gu, Y. Ando, Y. Tokura, S. Uchida, and 
N. P. Ong, Science \textbf{299}, 86 (2003).

\bibitem{WangNew}
Y. Wang, S. Ono, S. Komiya, Y. Ando, and N. P. Ong, 
(unpublished).

\bibitem{QPT}
Y. Ando, S. Ono, X. F. Sun, J. Takeya, F. F. Balakirev, J. B. Betts, 
and G. S. Boebinger, Phys. Rev. Lett. \textbf{92}, 247004 (2004).

\bibitem{Hanaki}
Y. Ando, Y. Hanaki, S. Ono, T. Murayama, K. Segawa, N. Miyamoto, 
and S. Komiya, Phys. Rev. B \textbf{61}, R14956 (2000); 
\textbf{63}, 069902(E) (2001).

\bibitem{logT}
Y. Ando, G. S. Boebinger, A. Passner, T. Kimura, and K. Kishio, 
Phys. Rev. Lett. \textbf{75}, 4662 (1995). 

\bibitem{MI}
G. S. Boebinger, Y. Ando, A. Passner, T. Kimura, M. Okuya, J. Shimoyama,
 K. Kishio, K. Tamasaku, N. Ichikawa, and S. Uchida, 
Phys. Rev. Lett. \textbf{77}, 5417 (1996).

\bibitem{OnoMI}
S. Ono, Y. Ando, T. Murayama, F. F. Balakirev, J. B. Betts, and 
G. S. Boebinger, Phys Rev. Lett. \textbf{85}, 638 (2000).

\bibitem{note}
The most overdoped BSLCO sample reported in Ref. \onlinecite{WangHc2} 
was $p$ = 0.18 ($x$ = 0.23), but it was incorrectly shown in that 
paper as $p$ = 0.21.

\bibitem{Bergemann}
C. Bergemann, A. W. Tyler, A. P. Mackenzie, J. R. Cooper, S. R. Jullian, and
D. E. Farrell, Phys. Rev. B \textbf{57}, 14387 (1998).

\bibitem{note2}
Note that this putative $\rho_c^n$ in the absence of the pseudogap is
different from the $\rho_c$ in the ``fully-resistive" state which is
realized in the presence of the pseudogap.

\bibitem{Eisaki}
H. Eisaki, N. Kaneko, D. L. Feng, A. Damascelli, P. K. Mang, K. M. Shen, 
Z.-X. Shen, and M. Greven, Phys. Rev. B \textbf{69}, 064512 (2004).

\end{thebibliography}
\end{document}